\documentclass[11pt]{article}
\usepackage[latin1]{inputenc}
\usepackage[english]{babel}
\usepackage[namelimits]{amsmath}
\usepackage{amssymb}
\usepackage{amsmath}
\usepackage{amsthm}

\begin{document}

\title{A technique for generating new solutions of Einstein's equations}

\author{Roberto Bergamini (1940-2003)
\\
Istituto di radioastronomia I.R.A., C.N.R., Bologna-Italy,\\
Stefano Viaggiu
\\
Dipartimento di Fisica Universit\'a di Bologna-Italy\\
E-mail:stefano.viaggiu@ax0rm1.roma1.infn.it}
\date{\today}\maketitle

\begin{abstract}
We present a simple technique for generating new solutions of Einstein's equations
using such function transformations
that leave the field equations in the Ernst form. In this context we
recover all the known covariant transformations of Ernst equations and we find the role
of the analytic ones. Finally we obtain a new asymptotically flat solution
starting from the Kerr solutions.
\end{abstract}	

\section{Introduction}

An interesting class of solutions in general relativity are those
stationary and axisymmetric. In particular the 
ones which are asymptotically flat are candidate for the description
of the exterior gravitational field of a uniformly
rotating star. The only known stationary axisymmetric asymptotically 
flat solutions are the Kerr solutions \cite {Kerr},
the generalized Tomimatsu solutions \cite{TS} and the Papapetrou solutions
\cite{Pap}. One would like, therefore, to obtain more such solutions.
Ehlers first \cite{JE} showed how it is possible to construct new 
stationary exterior solutions and
stationary interior solutions of the Einstein's field equation
starting from static exterior solutions by applying certain conformal
transformations to auxiliary metrics defined on 
three-dimensional manifolds in space-time. These are one-parameter
family of solutions. Geroch \cite{GR} has shown that one can, in fact, obtain
an infinite-parameter family. Further Xanthopoulos \cite{Xan} has given
a technique for generating from any one-parameter family of vacuum solutions of Einstein's
equation a two-parameter family. In this paper we introduce a simple 
but very powerful method for generating an infinite
tower of solutions. We start from a known solution using function transformations
on the complex potential $\epsilon=f+\imath \Phi$ which leaves the 
Ernst equations covariants in form. The plane of this paper is the
following. In section 2 we make an overview on the Ehlers method for
vacuum solutions in relation with the Papapetrou gauge for stationary 
axisymmetric space-time and we find a functional relation between $f$ and $\Phi$.
In sections 3-5 we derive the basic equations and in section 7
these are appropriately reduced. In sections 6,8-9
we integrate the basic equations in simple cases. Finally,
in section 10 we derive a new asymptotically flat solution
starting from the Kerr solutions. 

\section{Ehlers method for the vacuum}

The Ehlers method \cite{JE} makes possible to construct new stationary
exterior and interior solutions of the field equations of general
relativity starting from static solutions. Ehlers method for the vacuum affirms that
if $u=u_idx^i$ is a linear differential form and we  take a solution of the
static form
\begin{equation}
G=e^{-2U}H-e^{2U}dt^2\;,\;{\nabla}^2U=0,
\label{ax0}
\end{equation}
where $H=h_{ik}dx^idx^k$ is a quadratic differential form, is then
possible to obtain solutions of the form
\begin{equation}
G=\cosh(2U)H-{\cosh}^{-1}(2U){(dt-u)}^2,
\label{ax1}
\end{equation}
provided that the following condition is satisfied
\begin{equation}
{\eta}_{jkl}U^{,l}=u_{[j,k]},
\label{ax2}
\end{equation}
where ${\eta}_{jkl}$ denotes the usual totally skew-symmetric tensor
with components $0,\pm\sqrt{h}\;,\;h=|h_{ik}|$ and $'',''$ denotes the ordinary
derivate
. In relation to equation (\ref{ax1})
for the vacuum axisymmetric case we have, in Papapetrou gauge \cite{Pap},
$u_1=u_2=0\;,\;u_3=\omega$, with
\begin{equation}
H=e^{2\gamma}[{(dx^1})^2+{(dx^2)}^2]+{\rho}^2{(dx^3)}^2.
\label{ax3}
\end{equation}
Starting from the Ernst equations \cite{Ernst} with 
$f^{-1}=\cosh(2U)$ and with the equation for 
$\omega$ given by
\begin{equation}
{\omega}_{AA}-{\omega}_A\frac{{\rho}_A}{\rho}+\frac{2}{f}
{\omega}_Af_A=0\;,\;A=1,2,
\label{ax4}
\end{equation}
it is easy to show that the solution of (\ref{ax4}) and (\ref{ax2}) is given by
\begin{equation}
{\omega}_1=2\rho U_2\;,\;{\omega}_2=-2\rho U_1. 
\label{ax5}
\end{equation}
From equation (\ref{ax5}) it follows that ${\nabla}^2U=0$. For the metric function 
$f$ we have
\begin{equation}
2U_A=-\frac{f_A}{f\sqrt{1-f^2}}.
\label{ax6}
\end{equation}
From the Ernst potential $\Phi$ given by
\begin{eqnarray}
{\omega}_1=-\frac{\rho}{f^2}{\Phi}_2\;,\;
{\omega}_2=\frac{\rho}{f^2}{\Phi}_1,\label{ax7}
\end{eqnarray}
it follows that
\begin{equation}
{\Phi}_A=-\frac{ff_A}{\sqrt{1-f^2}}
\label{ax8}
\end{equation}
and therefore
\begin{equation}
\Phi=\sqrt{1-f^2}+c.
\label{ax99}
\end{equation}
The constant $c$ is inessential. Finally we find
\begin{equation}
f^2+{\Phi}^2=1.
\label{ax9}
\end{equation}
The condition (\ref{ax9}) caracterizes the vacuum stationary axisymmetric
solutions achievable with the Ehlers method compatible with the
Papapetrou gauge. The condition (\ref{ax9}) also caracterizes
the Papapetrou class of solutions \cite{Pap}. In Appendix A we give
some solutions derived from the relation (\ref{ax9}).
In section 8 we see how a ``gauge'' version of the condition
(\ref{ax9}) emerges in the context of our method, that can be
therefore considered a generalization of the Ehlers method for the vacuum case.
It is easy to show that the condition (\ref{ax9}) leads to solutions which
are the imaginary part of the Ernst $\xi$ function with
$\epsilon =f+\imath\Phi=\frac{\xi-1}{\xi+1}$. 
\section{Some basic facts}

If we take the Ernst equations \cite{Ernst}
\begin{eqnarray}
& &f{\nabla}^2f+{\Phi}_{A}^2-f_{A}^2=0, \nonumber\\
& &f{\nabla}^2\Phi-2f_A{\Phi}_A=0 \label{e100}
\end{eqnarray}
and if we perform, based on (\ref{e100}), a transformation in the
new variables $a$ and $b$ such that
\begin{eqnarray}
& &a{\nabla}^2a+b_{A}^2-a_{A}^2=0, \nonumber\\
& &a{\nabla}^2b-2a_A b_A=0, \label{e104}
\end{eqnarray}
then the transformation $f=f(a,b)$, $\Phi=\Phi(a,b)$ maps old solutions 
$f$ and $\Phi$ in the new solutions
$a$ and $b$, where $a$ and $b$ are the new $f$ and $\Phi$, 
respectively. Viceversa, if $a$ and $b$ are old solutions,
$f$ and $\Phi$ are new solutions. 
These transformations provide significantly new solutions
derived from the circumstance that an integration is involved in passing from $\Phi$
to the metric functions.

\section{Formulation of the problem}
Beginning from (\ref{e100}) we pose $f=f(a,b)$ and $\Phi=\Phi(a,b)$, 
from which, after a simple but very tedious algebra, 
(where index $A=1,2$ corresponds to the spatial derivates with 
respect to $x^1$ and $x^2$ 
and index $a$ and $b$ denote the derivate with respect to  
the variables $a$ and $b$) it follows
\begin{eqnarray}
& &fJ{\nabla}^2a+La_{A}^2+Mb_{A}^2+2Na_A b_A=0,\label{e114}\\
& &fJ{\nabla}^2b+\tilde{L}a_{A}^2+\tilde{M}b_{A}^2+
2\tilde{N}a_A b_A=0,\label{e115}
\end{eqnarray}
where $J=f_a{\Phi}_b-f_b{\Phi}_a$ and
\begin{eqnarray}
& &L=f(f_{aa}{\Phi}_b-{\Phi}_{aa}f_b)+{\Phi}_b({\Phi}_{a}^2-f_{a}^2)+
2f_a f_b{\Phi}_a, \label{e123}\\
& &M=f(f_{bb}{\Phi}_b-{\Phi}_{bb}f_b)+{\Phi}_b({\Phi}_{b}^2+f_{b}^2),\nonumber\\
& &N=f(f_{ab}{\Phi}_b-{\Phi}_{ab}f_b)+{\Phi}_a(f_{b}^2+{\Phi}_{b}^2),\nonumber\\
& &\tilde{L}=-f(f_{aa}{\Phi}_a-{\Phi}_{aa}f_a)-{\Phi}_a(f_{a}^2+{\Phi}_{a}^2),\nonumber\\
& &\tilde{M}=-f(f_{bb}{\Phi}_a-{\Phi}_{bb}f_a)-{\Phi}_a
({\Phi}_{b}^2-f_{b}^2)-2f_a f_b{\Phi}_b, \nonumber\\
& &\tilde{N}=-f(f_{ab}{\Phi}_a-{\Phi}_{ab}f_a)-
{\Phi}_b(f_{a}^2+{\Phi}_{a}^2).\nonumber
\end{eqnarray}
Note that
\begin{equation}
fJ_a={(f_a{\Phi}_b-f_b{\Phi}_a)}_a=
f(f_{aa}{\Phi}_b-f_b{\Phi}_{aa})-f(f_{ab}{\Phi}_a-{\Phi}_{ab}f_a),
\nonumber
\end{equation}
but from equation (\ref{e123}) we find
\begin{equation}
L+\tilde{N}=f^3{\left(\frac{J}{f^2}\right)}_a.
\label{e129}
\end{equation}
In the same way from $fJ_b=f{(f_a{\Phi}_b-f_b{\Phi}_a)}_b$
we obtain
\begin{equation}
N+\tilde{M}=f^3{\left(\frac{J}{f^2}\right)}_b.
\label{e130}
\end{equation}

\section{Basic equations}

We define
\begin{equation}
X=\frac{1}{f}f_a\;,\;Y=\frac{1}{f}f_b\;,\;
\xi=\frac{1}{f}{\Phi}_a\;,\;\eta=\frac{1}{f}{\Phi}_b
\label{e200}
\end{equation}
and then 
\begin{eqnarray}
& &L=f^3[\eta X_a-Y{\xi}_a+\eta{\xi}^2+XY\xi],\nonumber\\
& &M=f^3[\eta Y_b-Y{\eta}_b+\eta({\eta}^2+Y^2)],\nonumber\\
& &N=\frac{f^3}{2}[\eta(X_b+Y_a)-Y({\xi}_b+{\eta}_a)+
\xi({\eta}^2+Y^2)+\eta(XY+\eta\xi)],\nonumber\\
& &\tilde{L}=-f^3[\xi X_a-X{\xi}_a+\xi(x^2+{\xi}^2)],\nonumber\\
& &\tilde{M}=-f^3[\xi Y_b-X{\eta}_b+\eta(XY+\xi\eta)],\nonumber\\
& &\tilde{N}=-\frac{f^3}{2}[\xi(X_b+Y_a)-X({\xi}_b+{\eta}_a)+
\eta(X^2+{\xi}^2)+\xi(XY+\xi\eta)].\nonumber
\end{eqnarray}
From equation (\ref{e100}) taking $f=f(a,b)$ and $\Phi=\Phi(a,b)$
we obtain equations (\ref{e114})-(\ref{e115}) with
\begin{equation}
J=f^2(X\eta-Y\xi)\neq 0.
\label{e207}
\end{equation}
For the covariance to arise and consequently for equations (\ref{e114})-(\ref{e115})
to take the form (\ref{e104}), first we must have
\begin{equation}
\tilde{L}=\tilde{M}=N=0.
\label{e208}
\end{equation}
Besides we have equations (\ref{e129})-(\ref{e130}) with the condition
\begin{equation}
-L=M=-\tilde{N}
\label{e209}
\end{equation}
and therefore equation (\ref{e129}) becomes
\begin{equation}
2\tilde{N}=f^3{\left(\frac{J}{f^2}\right)}_a,
\label{e210}
\end{equation}
with
\begin{equation}
L=-M=\frac{f^3}{2}{\left(\frac{J}{f^2}\right)}_a.
\label{e211}
\end{equation}
While for equation (\ref{e130}) we get
\begin{equation}
N=f^3{\left(\frac{J}{f^2}\right)}_b
\label{e212}
\end{equation}
and then
\begin{eqnarray}
& &f^3\frac{J}{f^2}{\nabla}^2a+\frac{f^3}{2}{\left(\frac{J}{f^2}\right)}_a
\left(a_{A}^2-b_{A}^2\right)+f^3{\left(\frac{J}{f^2}\right)}_b a_A b_A=0,\label{e213}\\
& &f^3\frac{J}{f^2}{\nabla}^2b-\frac{f^3}{2}
{\left(\frac{J}{f^2}\right)}_b\left(a_{A}^2-b_{A}^2\right)+
f^3{\left(\frac{J}{f^2}\right)}_a a_A b_A=0 \label{e214}
\end{eqnarray}
and finally
\begin{equation}
{\left(\frac{J}{f^2}\right)}_b=0\;\;,\;\;\frac{J}{f^2}=A(a)=\frac{1}{a^2}.
\label{e215}
\end{equation}
The equations (\ref{e213})-(\ref{e214}) are the same as equations (\ref{e104}) and therefore
covariance follows.

\section{Case with $\xi=\alpha X$}

At the first stage we consider the equation $\tilde{L}=0$
\begin{equation}
\xi X_a-X{\xi}_a+\xi(X^2+{\xi}^2)=0.
\label{e300}
\end{equation}
The equation (\ref{e300}) permits to take
\begin{equation}
\xi=\alpha X
\label{e301}
\end{equation}
and from equation (\ref{e300}) we have, after an integration
\begin{equation}
\alpha=\frac{\epsilon f}{\sqrt{B^2-f^2}}\;\;\;\;\;\epsilon=\pm 1.
\label{e305}
\end{equation}
Where $B=B(b)$ is an integration constant.
In the same way, taking equation (\ref{e305}) into (\ref{e301}), we obtain
\begin{equation}
\tilde{B}-\Phi=\epsilon\sqrt{B^2-f^2},
\label{e306}
\end{equation}
where $\tilde{B}$ also depends only by $b$. Deriving
equation (\ref{e306}) with respect to  $b$ we obtain
\begin{equation}
\eta=\frac{1}{f}\left[{\tilde{B}}_b-\epsilon BB_b\frac{1}{\sqrt{B^2-f^2}}\right]+\alpha Y.
\label{e308}
\end{equation}
The four variable $X,Y,\xi,\eta$ are then related by
\begin{equation}
\xi=\alpha X=\frac{f}{\tilde{B}-\Phi}X\;,\;
\eta=\frac{\Pi}{f}+\alpha Y=\frac{\Pi}{f}+\frac{f}{\tilde{B}-\Phi}Y,
\label{e310}
\end{equation}
where
\begin{equation}
\Pi={\tilde{B}}_b-BB_b\frac{1}{\tilde{B}-\Phi}\;,\;
{\Pi}^{\prime}=\tilde{B}{\tilde{B}}_b-\epsilon BB_b
-{\tilde{B}}_b\Phi. 
\label{e314}
\end{equation}
Beyond the six equations $L=-M=\tilde{N}=f^3/2{(J/f^2)}_a$ and 
$\tilde{L}=\tilde{M}=N=0$ we have condition (\ref{e207})
\begin{equation}
X\eta-Y\xi=A=\frac{1}{a^2},
\label{e315}
\end{equation}
but from equations (\ref{e310}) we have
\begin{equation}
\frac{A}{X}=\frac{\Pi}{f}\;\rightarrow\;X=\frac{Af}{\Pi}, 
\label{e316}
\end{equation}
from which it follows
\begin{equation}
-\frac{1}{f}\left[{\tilde{B}}_b-\frac{\epsilon}{B}B_b\sqrt{B^2-f^2}\right]=
-\frac{1}{a}-h(b).
\label{e318}
\end{equation}
We set $A^{\star}=1/a$ and $h(b)=B^{\star}$. 
In this case we can obtain $f$ as a function of
$A^{\star}$ , $B^{\star}$, ${\tilde{B}}_b$ and $B$. With a little algebra from
(\ref{e318}) we find
\begin{equation}
\mu f^2-2\nu f+\theta=0,
\label{e321}
\end{equation}
where
\begin{equation}
\mu={(A^{\star}+B^{\star})}^2+\frac{1}{B^2}B_{b}^2\;,\;
\nu={\tilde{B}}_b(A^{\star}+B^{\star})\;,\;
\theta={\tilde{B}}_{b}^2-B_{b}^2.
\label{e322}
\end{equation}
Finally from the equation (\ref{e321}) we obtain
\begin{equation}
f=\frac{\nu\pm\sqrt{{\nu}^2-\mu\theta}}{\mu}\;.
\label{e323}
\end{equation}
Note that we have obtained the equation (\ref{e323}) 
using only the equation $\tilde{L}=0$. We must therefore check this
solutions with the others equations. In the next section we reduce
the relevant equations. 

\section{Reduction of the relevants equations}

We start from the six equations quoted above
\begin{eqnarray}
& &\eta(X_a+{\xi}^2)-Y({\xi}_a-X\xi)=\frac{1}{2}A_a, \label{e400}\\
& &\eta(Y_b+{\eta}^2)-Y({\eta}_b-Y\eta)=-\frac{1}{2}A_a, \label{e401}\\
& &\eta[(X_b+Y_a)+2\xi\eta]-Y[({\xi}_b+{\eta}_a)-(X\eta+Y\xi)]=0,\label{e402}\\
& &\xi(X_a+{\xi}^2)-Y({\xi}_a-X\xi)=0,\label{e403}\\
& &\xi(Y_b+{\eta}^2)-X({\eta}_b-Y\eta)=0,\label{e404}\\
& &\xi[(X_b+Y_a)+2\xi\eta]-X[{\xi}_b+{\eta}_a-(X\eta+Y\xi)]=-A_a,\label{e405}
\end{eqnarray}
with condition (\ref{e315}).
But $f(a,b)$, $\Phi(a,b)$ are not arbitrary. The following condition must be imposed
\begin{equation}
f_{ba}=f_{ab}\;\;,\;\;{\Phi}_{ab}={\Phi}_{ba},
\label{e407}
\end{equation}
from which it follows
\begin{equation}
X_b=Y_a\;,\;
{\xi}_b={\eta}_a+A.
\label{e411}
\end{equation}
Besides, differentiating condition  (\ref{e315}) we have
\begin{eqnarray}
& &\eta X_a-Y{\xi}_a-[\xi Y_a-X{\eta}_a]=A_a,\label{e412}\\
& &\eta X_b-Y{\xi}_b=\xi Y_b-X{\eta}_b.\label{e413}
\end{eqnarray}
Using equations (\ref{e411}), equation (\ref{e402}) becomes
\begin{equation}
2[(\eta Y_b-Y{\xi}_b)+\xi{\eta}^2+\eta XY]=0
\label{e414}
\end{equation}
and thanks to equation (\ref{e413}), equation (\ref{e414}) is the same as (\ref{e404}).
Similarly equation (\ref{e411}) placed into (\ref{e405}) leads to
\begin{eqnarray}
& &2[\xi Y_a-X{\eta}_a]+2{\xi}^2\eta-XA+X(X\eta+Y\xi)=\nonumber\\
& &=2[\xi Y_a-X{\eta}_a+2{\xi}^2\eta+2XY\xi]=\nonumber\\
& &=2[\eta X_a-Y{\xi}_a-A_a+\eta{\xi}^2+XY\xi]\Longrightarrow \nonumber\\
& &\eta(X_a+{\xi}^2)-Y({\xi}_a-X\xi)=\frac{1}{2}A_a. \label{e415}
\end{eqnarray}
But the equation (\ref{e415}) is the same as 
the (\ref{e400}). Therefore in place of equations 
 (\ref{e400})-(\ref{e405}) we have 
\begin{eqnarray}
& &\eta(X_a+{\xi}^2)-Y({\xi}_a-X\xi)=\frac{1}{2}A_a,\label{e416}\\
& &\eta(Y_b+{\eta}^2)-Y({\eta}_b-Y\eta)=-\frac{1}{2}A_a,\label{e417}\\
& &\xi(X_a+{\xi}^2)-X({\xi}_a-X\xi)=0,\label{e418}\\
& &\xi(Y_b+{\eta}^2)-X({\eta}_b-Y\eta)=0,\label{e419}
\end{eqnarray}
with the condition (\ref{e315}).
Thus the equations for the variable $a$ and $b$ are the
Ernst ones if and only if equations (\ref{e416})-(\ref{e419})
are satisfied. Note that this system is a first order system in the derivate 
$a$ and $b$, while the Ernst equations are a second order system in the 
coordinate derivates. In the next section we conclude the integration in the
case with $\xi=\alpha X$.

\section{Final solution in the case with  $\xi=\alpha X $}

At this stage we have used equation (\ref{e418}) only.
From the second equation of (\ref{e310}), with the help of (\ref{e301}),(\ref{e305}),
we get
\begin{equation}
Y=\eta\left(\frac{\tilde{B}-\Phi}{f}\right)-\frac{A}{\xi},
\label{e420}
\end{equation}
from which, after a derivation, we obtain
\begin{equation}
\xi(Y_b+{\eta}^2)-X({\eta}_b-Y\eta)=\xi
\left[\frac{\eta}{f}{\tilde{B}}_b-{\left(\frac{A}{\xi}\right)}_b\right].
\label{e425}
\end{equation}
If equation (\ref{e419}) is true, then it follows ($\xi\neq 0$)
\begin{equation}
\frac{\eta}{f}{\tilde{B}}_b={\left(\frac{A}{\xi}\right)}_b.
\label{e426}
\end{equation}
We suppose first that equation (\ref{e426}) is verified, then
equations (\ref{e416})-(\ref{e417}) become
\begin{eqnarray}
& &\left(\eta\frac{X}{\xi}-Y\right)\left({\xi}_a-X\xi\right)=
\frac{1}{2}A_a,\label{e427}\\
& &\left(\eta\frac{X}{\xi}-Y\right)\left({\eta}_b-Y\eta\right)=
-\frac{1}{2}A_a.\label{e428}
\end{eqnarray}
Equation (\ref{e427}) gives
\begin{equation}
\frac{\xi}{f\sqrt{A}}=B_o=B_o(b).
\label{e430}
\end{equation}
In relation to equation (\ref{e323})
and thanks to equation (\ref{e314}),(\ref{e316}) after a little algebra, we find
\begin{equation}
{\Pi}^{\prime}=-\epsilon f\sqrt{B^2{(A^{\star}+B^{\star})}^2-\theta}=
\frac{A^{\star}f}{B_o},
\label{e435}
\end{equation}
from which we get
\begin{equation}
\frac{{(A^{\star})}^2}{B_{o}^2}=B^2{(A^{\star}+B^{\star})}^2-\theta,
\label{e436}
\end{equation}
with the only solution given by
\begin{equation}
B^{\star}=0\; , \;\theta=0\; , \;B_o=\pm\frac{1}{B}.
\label{e437}
\end{equation}
We must consider the last two equations. At this stage we have the equations
\begin{equation}
f^2+{\Phi}^2=2B\Phi\epsilon\;,\;
f=\frac{\frac{2}{a}B_b}{\frac{1}{a^2}+\frac{B_{b}^2}{B^2}}.
\label{e439}
\end{equation}
Note that the condition (\ref{e439}) for $f$ and $\Phi$ 
can be considered a ``gauge'' version  of the
Ehlers condition (\ref{ax9}). The remaining equations are  the equation
(\ref{e426}) and (\ref{e428}). For equation (\ref{e428}) we have
\begin{equation}
\frac{\eta}{\xi}{\left[\ln\frac{\eta}{f}\right]}_b=-\frac{1}{2}\frac{A_a}{A}.
\label{e440}
\end{equation}
From the relation (\ref{e430}), after an integration, we find
\begin{equation}
\frac{\eta}{f}=\frac{G(b)+H(a)}{a^2},
\label{e444}
\end{equation}
where $G=\int\frac{1}{B}db$ and $H=H(a)$ is an arbitrary integration function. 
Placing equation (\ref{e444}) into (\ref{e426}) we obtain
\begin{equation}
[G+H]B_b={\left(\frac{1}{\xi}\right)}_b,
\label{e445}
\end{equation}
also, from (\ref{e430}), we have
\begin{equation}
{\left(\frac{1}{\xi}\right)}_b=-\frac{a}{f^2}\left[f_b B-fB_b\right].
\label{e446}
\end{equation}
With the expression for $f$ given by  (\ref{e439}),
equation (\ref{e445}) gives
\begin{eqnarray}
& &[G+H]B_b=\nonumber\\
& &=+\frac{a^2}{2B_{b}^2}\left[\frac{1}{a^2}(B_{b}^2-BB_{bb})-\frac{B_{b}^4}{B^2}+
\frac{B_{bb}B_{b}^2}{B}\right].\label{e448}
\end{eqnarray}
Equation (\ref{e448}) has two solutions. The first is the one with $H=0$, from which
it follows that $B$ is a constant, i.e. $f=0$, and therefore it is not acceptable.
The other solution is given by
\begin{equation}
GB_b=\frac{1}{2}\left(1-\frac{BB_{bb}}{B_{b}^2}\right)\;,\;
H=\epsilon a^2\;,\;B_b=\frac{1}{2}\frac{BB_{bb}-B_{b}^2}{B^2},
\label{e452}
\end{equation}
from which it follows
\begin{equation}
-\frac{1}{B}=2\epsilon b+c.
\label{e455}
\end{equation}
The constant $c$ is inessential. Therefore
we can choose $c=0$; in this case $G=-\epsilon b^2$. We get
\begin{equation}
f=\frac{a}{a^2+b^2}\;,\;
\frac{1}{f^2}{\Phi}_b=-\epsilon\frac{b^2}{a^2}+{\epsilon}^{\prime}
\;,\;{\epsilon}^{\prime}=\pm 1.
\label{e457}
\end{equation}
From the first equation of (\ref{e439}) we have
\begin{eqnarray}
& &{\Phi}^2+\epsilon\frac{\Phi}{b}+f^2=0 \Longrightarrow\nonumber\\
& &{\Phi}_1=-\frac{b}{a^2+b^2}\;,\;{\Phi}_2=-\frac{a^2}{b(a^2+b^2)}.
\label{e458}
\end{eqnarray}
Taking now the solution for  ${\Phi}_1$ and placing into (\ref{e457}) we have
\begin{equation}
\frac{b^2-a^2}{a^2}=\frac{-b^2\epsilon+{\epsilon}^{\prime}a^2}{a^2},
\label{e459}
\end{equation}
from which taking $\epsilon=-1$ and ${\epsilon}^{\prime}=-1$, 
it follows that the equation (\ref{e457}) is verified.
It is easy to show that the equation 
$\xi=(1/f){\Phi}_a$ is verified.
In the same way we can see that the solution ${\Phi}_2$ does not verify the condition
(\ref{e457}).
Then we have the final solution
\begin{equation}
f=\frac{a}{a^2+b^2}\;\;\;\Phi=-\frac{b}{a^2+b^2}.
\label{e460}
\end{equation}
The solution (\ref{e460}) represents the inverse transformation, i.e. it says that if
$\epsilon=f+\imath\Phi$ is a solution, also  $1/\epsilon$ is a solution. Although this is a known
transformation, it has been obtained with a new method.  
Our starting point is equation (\ref{e418}) with the assumption $\xi=\alpha X$. If we consider
a more general situation, i.e. $\xi=F(X,Y,\eta)$, we can obtain more general
solutions. Therefore a generalization of condition (\ref{e439}) appears. In this sense
our method can be considered a generalization of the Ehlers method.
In the case with $\xi=0$ it is easy to show that the only solution is the 
transformation
$f=a\;,\;\Phi=b+const.$ which  is the identity transformation. It is reassuring that
if we consider the most simple ansatz we obtain the most simple solution.
The search of new transformations is an open question,
but our method gives a tool in such direction.
In the next section we consider the analytic case, i.e. $F=-Y$.

\section{Analytic case}

In the analytic case we have for the variable  $X, \eta, Y, \xi$ the conditions
\begin{equation}
X=\eta\;\;\;,\;\;\;Y=-\xi,
\label{e600}
\end{equation}
with 
\begin{equation}
X\eta-Y\xi=\frac{1}{a^2}\Longrightarrow X^2+Y^2=\frac{1}{a^2}.
\label{e6000}
\end{equation}
The main equations in this case are
\begin{eqnarray}
& &X(X_a+Y^2)-Y(XY-Y_a)=-\frac{1}{a^3},\label{e601}\\
& &X(Y_b+X^2)-Y(X_b-XY)=\frac{1}{a^3},\label{e602}\\
& &XY_a-YX_a-Y(X^2+Y^2)=0,\label{e603}\\
& &-Y(Y_b+X^2)=X(X_b-XY).\label{e604}
\end{eqnarray}
At first we check the  consistency of our equations.
From equation (\ref{e604}) we have
\begin{equation}
X_b-XY=-\frac{Y}{X}\left(Y_b+X^2\right),
\label{e605}
\end{equation}
which taken into (\ref{e602}) and using equation (\ref{e6000}) gives
\begin{equation}
Y_b+X^2=\frac{X}{a}.
\label{e606}
\end{equation}
Equation (\ref{e601}) is expression (\ref{e6000}) differentiated with respect to $a$.                  
From equation  (\ref{e6000}) we have
\begin{equation}
X=\frac{1}{a}\sqrt{1-a^2Y^2}.
\label{e609}
\end{equation}
In the expression  (\ref{e609}) we have choosen the plus sign.
We have the conditions (\ref{e411}) from which it follows
\begin{equation}
Y_b=-X_a-\frac{1}{a^2}.
\label{e610}
\end{equation}
From equation (\ref{e603}) we obtain
\begin{equation}
X_a=\frac{X}{Y}Y_a-\frac{1}{a^2}
\label{e611}
\end{equation}
and then expression  (\ref{e610}) becomes
\begin{equation}
Y_b=-\frac{X}{Y}Y_a.
\label{e612}
\end{equation}
Differentiating expression (\ref{e609}) with respect to $a$,
thanks to (\ref{e612}), we have
\begin{equation}
X_a=-\frac{1}{a^2}\sqrt{1-a^2Y^2}-\frac{Y^2}{\sqrt{1-a^2Y^2}}+
\frac{a^2Y^2Y_b}{1-a^2Y^2},
\label{e613}
\end{equation}
which taken into (\ref{e610}) , after a little algebra, it gives
equation (\ref{e606}). Therefore our equations are consistent.
From  equation (\ref{e606}), after an integration, we get
\begin{equation}
Y^2[a^2+{(b+H(a))}^2]+2Y[b+H(a)]=0,
\label{e618}
\end{equation}
where $H=H(a)$ is an integration constant.
A simple solution is  $Y=0$, which leads to the identity transformation, a
obviously trivial analytic transformation. The other solution is
\begin{equation}
Y=-\frac{2[b+H(a)]}{[a^2+{(b+H(a))}^2]}
\label{e619}
\end{equation}
and then from  $f_b/f=Y$ we obtain
\begin{equation}
f=\frac{e^{G(a)}}{[a^2+{(b+H)}^2]}.
\label{e620a}
\end{equation}
We must finally consider the condition (\ref{e6000}) with $X=f_a/f$. We obtain
\begin{eqnarray}
& &G_{a}^2-\frac{1}{a^2}+\frac{4}{[a^2+{(b+H)}^2]}-
\frac{4aG_a}{[a^2+{(b+H)}^2]}-\nonumber\\
& &-\frac{4H_a(b+H)}{[a^2+{(b+H)}^2]}\left[\frac{H_a(b+H)}{[a^2+{(b+H)}^2]}+
G_a+a\right]=0.\label{e620}
\end{eqnarray}
Because $H$ and $G$ depend on the variable $a$, we conclude that the only solution
of the equation
(\ref{e620}) is obtained posing $G=\ln a$ and $H=c$, where $c$ is a constant.
If we choose $c=0$ we obtain again the transformation
\begin{equation}
f=\frac{a}{a^2+b^2}
\label{e621}
\end{equation}
and for the analyticity condition we must have
\begin{equation}
\Phi=-\frac{b}{a^2+b^2},
\label{e622}
\end{equation}
which represents again the inverse transformation. 
It is a simple matter to verify that choosing the minus sign in  expression
(\ref{e609}) we can never obtain an analytic solution.
We set now in equation (\ref{e620a}) $H=-c$ and therefore we obtain
\begin{equation}
f=\frac{a}{a^2+b^2+c^2-2bc}.
\label{e624}
\end{equation}
With simple manipulations (if $a$ and $b$ are solutions, also $ca$ and $cb$ are solutions) 
we see that the solution (\ref{e624}) can be written in  the following way
\begin{equation}
f=\frac{a}{c^2a^2+c^2b^2+1-2cb},
\label{e625}
\end{equation}
which for $\Phi$ it gives
\begin{equation}
\Phi=\frac{b-cb^2-ca^2}{1+c^2a^2+c^2b^2-2cb}.
\label{e626}
\end{equation}
The solutions (\ref{e625})-(\ref{e626}) say that if $\epsilon=a+\imath b$ is a solution,
also it is a solution the transformation
\begin{equation}
\epsilon\;\Longrightarrow\;\frac{\epsilon}{1+\imath c\epsilon}.
\label{e627}
\end{equation}
The solution (\ref{e627}) represents the so called Ehlers transformation. We can finally 
conclude with the claim that the only   analytic, not trivial, transformations 
which allow the covariance of the
Ernst equations, are the inverse transformations and the Ehlers transformations. 
Note that we can not obtain an infinite tower of solutions starting from
the inverse transformation, because the inverse transformation of the inverse one
is the identity transformation. For the
Ehlers transformation we have that, after $n$-transformations, 
$\epsilon\rightarrow\frac{\epsilon}{1+\imath cn\epsilon}$ 
. We can start from two arbitrary transformations
to obtain infinite combinations that allow infinite covariant transformations.
In the next section we apply the inverse transformation to the Kerr solutions

\section{Function transformations for the Kerr metric}

We consider for simplicity the case of the extreme Kerr solution \cite{Kerr}
expressed in spherical coordinates with $r=e^{v}$. We start from the line element
\begin{equation}
ds^2=f^{-1}[e^{2\gamma}(dv^2+d{\vartheta}^2)+{\rho}^2d{\varphi}^2]-
f{(dt-\omega d\varphi)}^2.
\label{acci}
\end{equation}
We have for this solution
\begin{eqnarray}
& &a=f_{Kerr}=\frac{p^2e^{2v}-{\sin}^2\vartheta}
{{(pe^{v}+1)}^2+{\cos}^2\vartheta},\label{e700}\\
& &b={\Phi}_{Kerr}=\frac{2\cos\vartheta}{{(pe^v+1)}^2+{\cos}^2\vartheta}.\label{e701}
\end{eqnarray}
Performing the inverse transformation we obtain for the new  $f$ and $\Phi$
the expression
\begin{eqnarray}
& &f=\frac{\left[p^2e^{2v}-{\sin}^2\vartheta\right]
\left[{(pe^v+1)}^2+{\cos}^2\vartheta\right]}
{{(p^2e^{2v}-{\sin}^2\vartheta)}^2+4{\cos}^2\vartheta},\nonumber\\
& &\Phi=-\frac{2\cos\vartheta\left[{(pe^v+1)}^2+{\cos}^2\vartheta\right]}
{{(p^2e^{2v}-{\sin}^2\vartheta)}^2+4{\cos}^2\vartheta}.\label{e702}
\end{eqnarray}
We considere now the asymptotical behaviour of the solution 
(\ref{e702}). First we note that for $v\rightarrow\infty$  $f\rightarrow 1$
and $\Phi\rightarrow 0$. Besides we have
\begin{equation}
{\Phi}_{\vartheta}\simeq \frac{2\sin\vartheta}{p^2e^{2v}}+O(e^{-3v})\;,\;
{\Phi}_v\simeq \frac{4\cos\vartheta}{p^2e^{2v}}+O(e^{-3v}),
\label{e703}
\end{equation}
from which, thanks to (\ref{ax7}), we get
\begin{equation}
{\omega} \simeq \frac{2}{p^2}{\sin}^2\vartheta e^{-v}+O(e^{-2v})
\label{e705}
\end{equation}
and therefore for $v\rightarrow\infty$ $\omega\rightarrow 0$. From the Ernst
equations \cite{Ernst} it follows that
\begin{equation}
\gamma \simeq v-\frac{1}{2p^2}{\sin}^2\vartheta e^{-2v}+O(e^{-3v}).
\label{e709}
\end{equation}
Setting $r=e^v$ we obtain
\begin{equation}
f\simeq 1+\frac{2}{pr}\;,\;
\omega \simeq \frac{2}{rp^2}{\sin}^2\vartheta\;,\;
\gamma \simeq \ln r-\frac{1}{2p^2 r^2}{\sin}^2\vartheta.
\label{e710}
\end{equation}
Therefore the solution (\ref{e702}), obtained performing the 
inverse transformation to the 
extreme Kerr solution, is asymptotically flat. 
Besides, thanks to expansion (\ref{e710}), the 
solution (\ref{e702}) represents a  metric with a spinning source
if and only if  $p=-1/m$, where $m$ is the mass of the source. 
Similar arguments follow for the other Kerr
solutions. The application of the Ehlers transformations to the Kerr metric
does not give asymptotically flat solutions. We conclude with a remark
on the Ehlers transformations. 
Consider now the transformation (\ref{e627}). We consider 
the case with $c=1$. Performing this transformation to the class
of solution with $f^2+{\Phi}^2=1$, we obtain
\begin{equation}
\epsilon =\frac{f}{2(1-\Phi)}-\frac{\imath}{2}.
\label{e712}
\end{equation}
Because the constant $c$ is an inessential constant 
(for $c\neq 1$, if $\epsilon$ is a solution also
$\frac{\epsilon}{c}$ is a solution)
we can say that performing 
a Ehlers transformation to the class of solutions given by 
the condition (\ref{ax9}) ($c=1$ or $\epsilon\rightarrow\frac{\epsilon}{c}$)
we obtain a static solution (up to a constant factor which can be 
set to 0).

\section{Conclusions}

In this paper we introduce a new, simple  
but very powerful method for generating
a tower of solutions starting from known ones using
the covariance of the Ernst equations for
generic metric functions transformations. 
The basic equations are derived and reduced. We have considered
particular cases obtaining all the transformations known in
literature. Finally, we obtain an asymptotically flat
solution starting from the Kerr one.

\section*{Acknowledgments}

I would like to thank Roberto Balbinot for his encouragement and many useful 
discussions.

\section*{Appendix A}

In relation to (\ref{ax9}) we consider first the 
cylindrical solution $\Phi=\Phi(\rho)$ and therefore $\omega=\omega(z)$.
This ansatz gives 
$f=\frac{2{\rho}^p}{1+{\rho}^{2p}}$ with $p$ 
real constant and integrating the 
field equations we have
\begin{eqnarray}
& &ds^2=\frac{{\rho}^{\frac{p^2}{2}}}{2}({\rho}^p+{\rho}^{-p})
(d{\rho}^2+dz^2)+\nonumber\\
& &+\frac{{\rho}^2}{2}({\rho}^p+{\rho}^{-p})d{\varphi}^2-
\frac{2{\rho}^p}{1+{\rho}^{2p}}{(dt-pzd\varphi)}^2.\label{ape1}
\end{eqnarray}
This is a cylindrical solution which is an anti-Machian solution.\\
It is possible to give a non cylindrical solution starting from the
equation (\ref{ape1}) using parabolic coordinates 
$\rho=2\lambda\mu,\;z={\lambda}^2-{\mu}^2$. The Laplacian in these
coordinates is
\begin{equation}
{\nabla}^2={\partial}_{\mu\mu}+{\partial}_{\lambda\lambda}+
\frac{{\partial}_{\lambda}}{\lambda}+\frac{{\partial}_{\mu}}{\mu},
\label{ape2}
\end{equation}
from which it follows that if the metric functions are indipendent from $\lambda$ (or $\mu$) 
the Laplacian (\ref{ape2}) is isometric to the cylindrical one.
Therefore if $f=f(\lambda)$ we obtain, similarly to the above cylindrical
solution,
\begin{eqnarray}
& &ds^2=\frac{({\lambda}^2+{\mu}^2)(1+{\lambda}^{2p})}
{2{\lambda}^p{\left(\frac{{\lambda}^2+{\mu}^2}
{{\lambda}^2}\right)}^{\frac{p^2}{4}}}
(d{\lambda}^2+d{\mu}^2)+\nonumber\\
& &+\frac{2{\mu}^2(1+{\lambda}^{2p})}{{\lambda}^{p-2}}d{\varphi}^2-
\frac{2{\lambda}^p}{1+{\lambda}^{2p}}
{(dt-p{\mu}^2d\varphi)}^2.
\label{ape3}
\end{eqnarray}
Equation (\ref{ape3}) has a complicated behaviour 
when expressed in  cylindrical coordinates.
If we change $\lambda$ with $\mu$ in the line element (\ref{ape3})
we have also a not cylindrical solution.
By using spherical coordinates
$U=-k/r$ is a solution of ${\nabla}^2U=0$. Therefore we have the spherical 
solution
\begin{eqnarray}
& &ds^2=e^{-\frac{k^2{\sin}^2\theta}{r^2}}\cosh(2k/r)[dr^2+r^2d{\theta}^2]+
r^2{\sin}^2\theta\cosh(2k/r)d{\varphi}^2-\nonumber\\
& &-\frac{1}{\cosh(2k/r)}{(dt-2k\cos\theta d\varphi)}^2.
\label{ape5}
\end{eqnarray}
Finally, taking $U=pz$ we obtain the solution
\begin{eqnarray}
& &ds^2=\cosh(2pz)e^{-p^2{\rho}^2}[d{\rho}^2+dz^2]+{\rho}^2\cosh(2pz)d{\varphi}^2
-\nonumber\\
& &-\frac{1}{\cosh(2pz)}{(dt-p{\rho}^2d\varphi)}^2.
\label{apein}
\end{eqnarray}
Note that taking $z=0$ in (\ref{apein}) we recover the Van Stokum
solutions \cite{VS} in a co-moving reference frame.

\end{document}